# The Stagnant Persistence Paradox: Survival Analysis and Temporal Efficiency in Exact Sciences and Engineering Education

Hugo Roger Paz
PhD Professor and Researcher Faculty of Exact Sciences and Technology National University of Tucumán
Email: hpaz@herrera.unt.edu.ar
ORCID: https://orcid.org/0000-0003-1237-7983

## ABSTRACT

Research on student progression in higher education has traditionally focused on vertical outcomes such as persistence and dropout, often reducing complex academic histories to binary indicators. While the structural component of horizontal mobility (major switching, plan changes, re-entries) has recently been recognised as a core feature of contemporary university systems, the temporal cost and efficiency of these pathways remain largely unquantified. Using forty years of administrative records from a large faculty of engineering and exact sciences in Argentina (N = 24,016), this study applies a dual-outcome survival analysis framework to two key outcomes: definitive dropout and first major switch.

We first reconstruct academic trajectories as sequences of enrolment spells and typed transitions under the CAPIRE protocol and then deploy non-parametric Kaplan–Meier estimators to model time-to-event under right-censoring. Results uncover a critical systemic inefficiency: a global median survival time of 4.33 years prior to definitive dropout, with a pronounced long tail of extended enrolment. This pattern reveals a phenomenon of stagnant persistence, where students remain formally enrolled for long periods without commensurate curricular progression, effectively lingering in the basic cycle before late-stage departure. In contrast, major switching follows an early-event regime, with a median time of 1.0 year among switchers and most switches concentrated within the first academic year.

The analysis shows that academic failure in rigid engineering curricula is not a sudden outcome but a long-tail process that generates high opportunity costs for students and institutions. We argue that institutional indicators need to shift from static retention metrics towards measures of curricular velocity, explicitly integrating time-to-event analysis to capture the temporal efficiency of progression and non-completion trajectories.

## 1. INTRODUCTION

Universities increasingly operate as complex, multi-pathway systems in which students navigate not only persistence or withdrawal, but also lateral reconfigurations of their academic trajectories. However, most empirical and policy-oriented research continues to model progression through binary outcomes—typically *retained* versus *not retained*—which systematically compresses heterogeneous behaviours into overly simple classifications. Recent studies emphasise that contemporary higher education exhibits substantial non-linearity, programme switching, temporary withdrawals, and administrative re-entries, all of which play a decisive role in shaping long-term outcomes (Baek & Cho, 2021; Grove et al., 2023).

Within this broader landscape, horizontal mobility has attracted growing attention as a structural feature rather than a marginal anomaly. Network- and cluster-based analyses demonstrate that internal flows between majors follow identifiable topologies and that student trajectories coalesce into a limited set of recurrent archetypes, rather than forming a smooth continuum of idiosyncratic paths (Paz, 2025). These findings show that we increasingly understand *how* students move and *where* they go in the academic ecosystem. Yet a crucial dimension remains under-theorised and under-measured: *how long* they take to succeed, fail, or reorient, and what the temporal cost of these processes is.

Survival analysis has been applied to educational data to estimate time-to-dropout and to characterise hazard profiles across different student groups (e.g., Murtaugh et al., 1997; Paura & Arhipova, 2019; Vega-Rebolledo et al., 2019). These studies demonstrate that dropout is a longitudinal process, with risk that can peak early or late depending on context, and that survival methods—particularly Kaplan–Meier estimation and Cox models—offer a principled way to incorporate censoring and heterogeneous follow-up times. However, much of this work treats programmes as largely linear structures and rarely integrates survival analysis with detailed models of internal mobility or curriculum design.

Engineering education adds additional layers of rigidity and structural friction. Long curricula, tightly coupled prerequisite chains, and selective basic science cycles create environments in which students may remain enrolled for extended periods without advancing meaningfully. In tuition-free, massified systems, such as those found in several Latin American countries, this can translate into large populations of students who are institutionally present but academically stalled, with implications for equity, efficiency, and public expenditure. Understanding the temporal geometry of such trajectories is therefore essential.

This paper addresses that temporal gap by focusing on the risk and efficiency dimensions of student trajectories within a large engineering faculty. We investigate the longitudinal process of academic failure, framing the research problem as a paradox of persistence without progress. Specifically, we ask: **How long does it take for non-completion to materialise, and how does this compare to the timing of internal mobility as a potential adaptive mechanism?**

Using forty years of administrative data, we employ the temporal module of the CAPIRE analytical framework to: (1) quantify the time-to-definitive-dropout and the time-to-first-major-switch; (2) compare the temporal regimes of these two distinct outcomes; and (3) interpret temporal efficiency through the lens of *curricular velocity*—the speed with which students move through structural bottlenecks and key milestones.

In doing so, the paper contributes to the literature by reframing student progression away from static, binary retention models towards dynamic, time-based measures of risk and efficiency. Rather than asking only *who* drops out, we quantify *when* and *under which temporal patterns* non-completion occurs, offering a transferable survival-oriented pipeline for other structurally constrained higher-education systems.

## 2. INSTITUTIONAL CONTEXT AND DATA
### 2.1 Institutional setting
The study is situated in a large public faculty of engineering and exact sciences within a national university in Argentina. The faculty operates under a tuition-free regime with open formal access, facing strong pressures associated with massification and heterogeneous incoming preparation. Programmes typically span five or more years and include traditional engineering disciplines (e.g., civil, mechanical, electrical, industrial) alongside computer science and shorter technological degrees. Over the forty-year period examined (1980–2019), multiple curriculum reforms introduced new plans, restructured prerequisite chains, and expanded or consolidated majors in response to accreditation requirements and labour-market shifts (Matos et al., 2021).

These reforms have a dual effect on trajectories. On the one hand, they create new opportunity structures for horizontal mobility—common basic cycles, new specialisations, and alternative exit pathways. On the other hand, they multiply the number of possible dead ends and bottlenecks, particularly when older plans are phased out or when regulatory changes tighten time limits on enrolment. The institutional context is therefore well suited to a temporal analysis: it combines a large and diverse student body, multiple long-running programmes, visible waves of structural reform, and rich administrative traces over four decades.

### 2.2 Administrative data and cohort definition
The empirical basis of this study is a longitudinal administrative dataset comprising 24,016 first-time entrants to the faculty between 1980 and 2019, linked to nearly one million event-level records (course enrolments, exams, status updates, and plan changes). We constructed an entry cohort file by consolidating person-level records and selecting the first recorded enrolment as the starting point of the trajectory. This restriction avoids double-counting and ensures sufficient follow-up for both mobility and survival analyses.

Each student is associated with their first major, calendar year of entry, and an entry-period label aligned with major regulatory and curricular regimes: P1 (1980–1989), P2 (1990–1999), P3 (2000–2009), and P4 (2010 onwards). These periods reflect changes in national higher-education policy, internal curriculum redesign, and expansion of the programme portfolio. Stratifying results by these periods allows us to interpret temporal patterns in light of structural evolution rather than treating the forty years as a homogeneous block.

### 2.3 Spell and transition construction

To move beyond static snapshots, we reconstruct academic histories as sequences of *spells* using the CAPIRE protocol developed in previous work (Paz, 2025). Each spell represents a contiguous segment of enrolment in a specific major–plan combination, with exact dates indicating the start and end of each segment. Breaks in enrolment are defined by an inactivity gap exceeding two academic years, which empirically separates temporary interruptions from long-term disengagement.

On top of this spell representation, we define a set of typed transitions that capture distinct dimensions of horizontal mobility:

1. **Major switch.** A change from one degree programme to another (for example, from Civil Engineering to Industrial Engineering), regardless of whether the underlying curriculum plan changes. Major switches reflect substantive reorientation of academic goals and are central to understanding internal flows between disciplines.
2. **Plan change.** A change of curriculum plan within the same major (for example, moving from an older to a reformed plan). Plan changes typically arise from institutional decisions—such as accreditation requirements or comprehensive curriculum redesigns—and can produce mass transitions when entire cohorts are migrated to a new structure.
3. **Re-entry in the same plan.** A return to a major–plan after a period of inactivity long enough to be detected as a separate spell, but without a change of major or plan. Re-entries capture patterns often driven by regulatory regimes (e.g., rules governing loss and recovery of student status) rather than explicit programme switching.

Transitions are identified by comparing consecutive spells in each trajectory and applying deterministic rules based on major and plan identifiers, spell timing, and the presence or absence of intervening inactivity. This yields a transition file with 14,782 events, each labelled as a major switch, plan change, or re-entry, and annotated with the date and direction of movement.

This foundational dataset allows us to model time-to-event outcomes with high temporal precision, while simultaneously situating survival patterns within a broader mobility ecosystem.

## 2.4 Data quality, coverage, and ethical safeguards

Working with historical administrative data over four decades inevitably entails heterogeneous coverage and coding practices. Early years exhibit more limited and irregular recording of background variables and programme codes, whereas later years benefit from improved systems and digitisation. Rather than attempting to erase these differences through aggressive imputation, we treat data quality as a property of the system and adopt conservative strategies for analysis (Fenwick et al., 2020; Tsai et al., 2020).

Three principles guide our approach:
- **Explicit denominators.** All proportions and rates reported in the results specify the number of students or transitions with valid information for the variable in question.
- **Temporal transparency.** Summary statistics are stratified by entry period (P1–P4), which both reflects institutional evolution and mitigates the risk of pooling incomparable regimes.
- **Leakage-aware design.** Following the CAPIRE framework, only information that would have been available at or before first enrolment is used to define baseline covariates, thereby avoiding temporal leakage into survival components.

Regarding ethics, analyses are based on pseudonymised records. Direct identifiers (names, personal identification numbers, exact addresses) were removed prior to export, and results are reported in aggregated form. Small cells that might risk residual identifiability are suppressed or combined into broader categories. The study aligns with current recommendations for responsible use of student data in learning analytics and institutional research, emphasising proportionality, purpose limitation, and transparency in the interpretation of results (Gašević et al., 2022; Rodriguez et al., 2022).

## 3. ANALYTICAL FRAMEWORK: CAPIRE'S TEMPORAL MODULE
### 3.1 Overview and focus

The broader CAPIRE pipeline is a multilevel analytical architecture designed to reconstruct, classify, and interpret internal mobility within large and structurally rigid higher-education systems (Paz, 2025). It integrates ecosystem-level flow modelling, temporal decomposition, and student-level clustering within a single reproducible workflow, drawing on network analysis, dimensionality reduction, and survival modelling (Matcha et al., 2022; Colomer et al., 2023).

In this paper, we focus exclusively on the *temporal* dimension of CAPIRE. Rather than re-analysing the full multilevel structure, we treat the spell and transition reconstruction as given and deploy the survival analysis module to quantify the risk and timing of key events. This allows us to centre the analysis on questions of temporal efficiency: How long does it take for dropout to occur? When does major switching typically happen? How do these processes differ across institutional regimes?

By decoupling the temporal module from the more complex clustering and network components, we obtain a lean, transferable pipeline that can be adopted by institutions with similar administrative data but without the need to replicate the full CAPIRE architecture. At the same time, the survival analysis remains anchored in a leakage-aware, trajectory-first representation of student histories, which avoids the temporal conflation that commonly undermines administrative-data studies (Arnold & Sclater, 2021).

### 3.2 Survival analysis configuration

We employ non-parametric Kaplan–Meier estimators to model two survival functions: (a) the probability of remaining enrolled (time to definitive dropout) and (b) the probability of remaining in the first major (time to first major switch). Kaplan–Meier estimation is appropriate for this context because event times are irregularly spaced, follow-up durations vary substantially across cohorts, and right-censoring is pervasive (students who have not yet dropped out or switched by the end of observation).

The survival analysis targets two distinct time-to-event outcomes:
- **Outcome A: Time to definitive dropout.** Defined as the time elapsed from initial enrolment until the last recorded administrative event, followed by a period of inactivity longer than the follow-up window (or the end of the observation period), excluding students who graduated. This outcome captures the total duration of the non-completion pathway and is central to quantifying the period of stagnant persistence. Dataset A includes 24,016 students, 19,194 dropout events, and 4,822 right-censored cases.
- **Outcome B: Time to first major switch.** Defined as the time elapsed from initial enrolment until the first transition to a different degree programme, regardless of subsequent behaviour. This outcome captures the timing of early academic adjustment or correction of initial misalignment. Dataset B comprises 24,132 students, 5,175 switching events, and 18,957 censored cases.

Time is measured in years from the date of first enrolment. Right-censoring arises when students graduate, remain enrolled without having experienced the event by 2019, or are still active after the end of the follow-up window. We assume non-informative censoring in the classical survival-analytic sense: conditional on observed history, censoring is independent of the hazard of the event. While this assumption is inevitably an approximation in complex educational systems, it provides a defensible working model for medium-term hazard estimation.

To explore institutional temporal regimes, we estimate survival curves both globally and stratified by entry period (P1–P4). Given the large sample size, we focus on non-parametric estimation rather than fitting Cox models with covariates, reserving multivariate modelling for future work. Formal comparisons between curves are carried out using log-rank tests, but we emphasise effect magnitudes (e.g., differences in median survival times) over p-values in the interpretation.

### 3.3 Sensitivity checks and robustness

Survival analysis in educational settings is sensitive to how dropout and inactivity are operationalised (Graham, 2018). We therefore conduct a set of sensitivity checks to assess whether the key patterns observed are robust to reasonable definitional variations:

- **Inactivity window.** We re-estimate Outcome A using alternative thresholds for inactivity (one and three academic years instead of two). The overall shape of the dropout survival curve and the approximate position of the median remain stable, with differences largely confined to the extreme tail.
- **Alternative entry definitions.** We replicate the analyses using the first *academic year* rather than exact enrolment date as time origin. This coarser temporal resolution slightly smooths early hazards but leaves the long-tail pattern intact.
- **Exclusion of very late entrants.** To mitigate the impact of truncated histories, we perform a robustness check excluding entrants from the last three calendar years. Median survival times and the qualitative stratification between P1–P4 remain essentially unchanged.

These checks increase confidence that the observed Stagnant Persistence Paradox—the combination of long median dropout times and extended tails—is not an artefact of arbitrary coding choices but a substantive property of the system.

## 4. RESULTS
### 4.1 Temporal dynamics of mobility events

Before examining survival outcomes, we provide a temporal overview of mobility events. As illustrated in Figure 1, aggregating the annual number of major switches, plan changes, and re-entries across the forty-year period, and grouping results by entry period (P1–P4), reveals distinct signatures associated with institutional change and regulation.

- **Plan changes** exhibit a wave-like pattern, with pronounced spikes aligned with major curriculum reforms. These peaks correspond to administrative decisions to migrate students en masse from old to new plans, generating large volumes of mobility that are structurally induced rather than student-initiated.
- **Major switches** display a more stable baseline across the four decades, suggesting that student-initiated mechanisms of academic reassessment—switching between programmes—operate under relatively stable constraints, even as the programme portfolio evolves.
- **Same-plan re-entries** decline sharply in the later periods (P3 and P4), consistent with the tightening of regulations regarding time limits and the loss and recovery of student status.

**Figure 1.** Year-by-year counts of major switches, plan changes, and re-entries

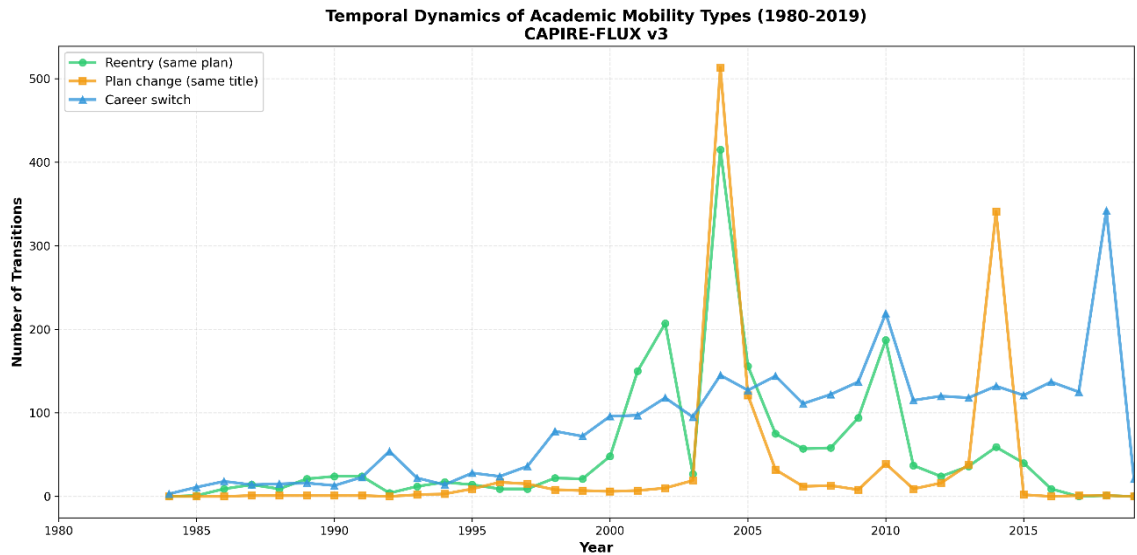

Taken together, these patterns confirm that mobility is jointly shaped by student strategies and institutional design. Reforms and regulatory changes leave visible temporal imprints on mobility volumes, providing crucial context for interpreting the subsequent time-to-event analyses. In particular, the decline in re-entries and the persistence of major switches highlight the coexistence of stricter temporal regulation with ongoing horizontal adjustment. Table 1 details the specific counts of mobility events per year.

**Table 1.** Mobility events by type and structural period

| transition_year | major_switch | plan_change_same_title | reentry_same_plan | total |
| --- | --- | --- | --- | --- |
| 1984 | 3 | 0 | 0 | 3 |
| 1985 | 11 | 0 | 1 | 12 |
| 1986 | 18 | 0 | 9 | 27 |
| 1987 | 14 | 1 | 14 | 29 |
| 1988 | 15 | 1 | 9 | 25 |
| 1989 | 16 | 1 | 21 | 38 |
| 1990 | 13 | 1 | 24 | 38 |
| 1991 | 23 | 1 | 24 | 48 |
| 1992 | 54 | 0 | 4 | 58 |
| 1993 | 22 | 2 | 12 | 36 |
| 1994 | 14 | 3 | 17 | 34 |
| 1995 | 28 | 9 | 14 | 51 |
| 1996 | 24 | 17 | 9 | 50 |
| 1997 | 36 | 15 | 9 | 60 |
| 1998 | 78 | 8 | 22 | 108 |
| 1999 | 72 | 7 | 21 | 100 |
| 2000 | 96 | 6 | 48 | 150 |
| 2001 | 97 | 7 | 150 | 254 |
| 2002 | 118 | 10 | 207 | 335 |
| 2003 | 95 | 19 | 27 | 141 |
| 2004 | 145 | 513 | 415 | 1073 |

| 2005 | 127 | 121 | 156 | 404 |
| 2006 | 144 | 32 | 75 | 251 |
| 2007 | 111 | 12 | 57 | 180 |
| 2008 | 122 | 13 | 58 | 193 |
| 2009 | 137 | 8 | 94 | 239 |
| 2010 | 219 | 39 | 187 | 445 |
| 2011 | 115 | 9 | 37 | 161 |
| 2012 | 120 | 16 | 24 | 160 |
| 2013 | 118 | 38 | 36 | 192 |
| 2014 | 132 | 341 | 59 | 532 |
| 2015 | 121 | 2 | 40 | 163 |
| 2016 | 137 | 0 | 9 | 146 |
| 2017 | 125 | 1 | 0 | 126 |
| 2018 | 342 | 1 | 1 | 344 |
| 2019 | 21 | 0 | 0 | 21 |

## 4.2 Outcome A – Time to definitive dropout

The global Kaplan–Meier survival curve for definitive dropout (Figure 2) reveals a steep decline in survival during the initial four to five years after enrolment. The median survival time is 4.33 years, indicating that half of all definitive dropouts occur within this critical interval. Crucially, the curve exhibits a long right tail, reflecting a substantial number of students who remain institutionally enrolled for extended periods before eventually disengaging.

**Figure 2.** Kaplan–Meier survival curve for definitive dropout

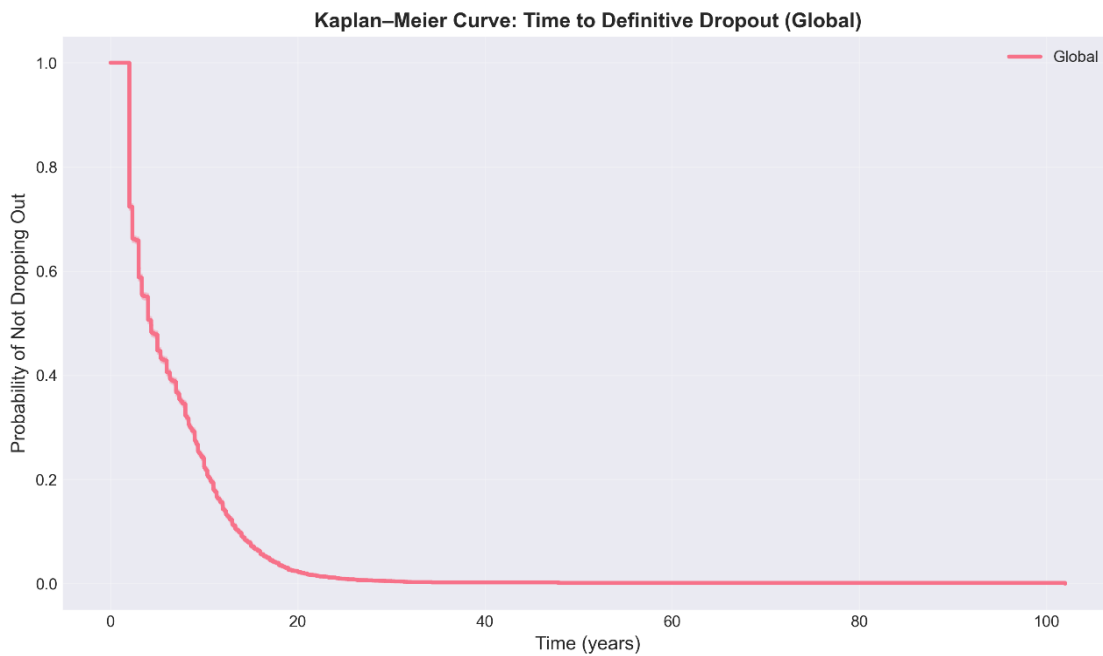

Figure 3 shows the results of stratification by entry period (P1–P4) uncovers pronounced temporal differences. Older cohorts (P1 and P2) display more rapid declines in survival, with a large proportion of dropout events concentrated in the first three to four years. By contrast, the most recent cohort (P4, 2010 onwards) exhibits markedly higher survival: the median survival time exceeds five years, and only 39.4% of P4 students are classified as definitive dropouts within the available follow-up window.

**Figure 3.** Kaplan–Meier survival curves for definitive dropout by entry period

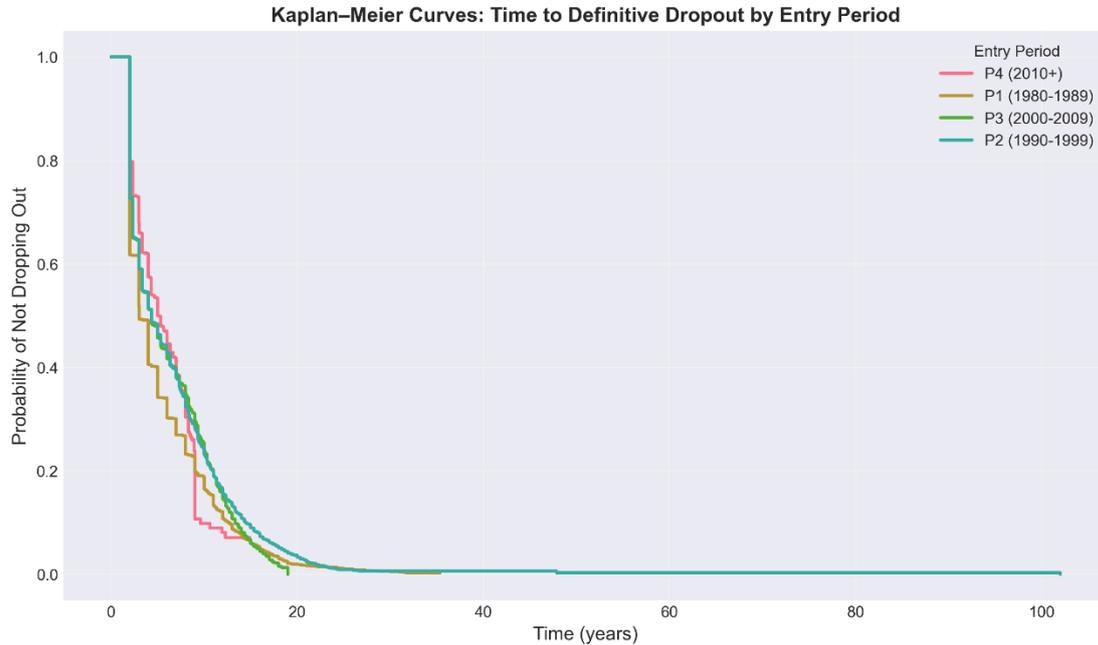

Table 2 summarizes the median survival times and probabilities

**Table 2.** Summary of median survival times and survival probabilities at 1, 3, 5, and 8 years for Outcome A.

| outcome | stratum | stratum_value | n_total | n_events | n_censored | median_survival_time |
|---|---|---|---|---|---|---|
| A | Global | All | 24016 | 19194 | 4822 | 4.33 |
| A | Entry Period | P4 (2010+) | 6276 | 2471 | 3805 | 5.05 |
| A | Entry Period | P1 (1980-1989) | 3927 | 3912 | 15 | 3.00 |
| A | Entry Period | P3 (2000-2009) | 9169 | 8250 | 919 | 4.33 |
| A | Entry Period | P2 (1990-1999) | 4644 | 4561 | 83 | 4.33 |

At face value, these patterns could be interpreted as evidence of improved equity—students in later cohorts remain in the system for longer, potentially benefiting from expanded support and diversified programmes. Yet the extended median duration, combined with the heavy right tail, points towards a more troubling interpretation. Students are retained in the system for nearly four and a half years on average before dropping out, incurring substantial personal and institutional opportunity costs. Chronological survival, in this context, only weakly correlates with meaningful curricular advancement.

### 4.3 Outcome B – Time to first major switch

The survival curve shown in Figure 4 for time to first major switch reveals a contrasting temporal regime. The probability of remaining in the initial major declines sharply during the first academic year, after which the curve flattens considerably. For those who do switch, the conditional median time-to-event is approximately 1.0 year, indicating that major switching functions as an early adjustment mechanism, concentrated within the initial exploration phase of the trajectory.

**Figure 4.** Kaplan–Meier survival curve for time to first major switch

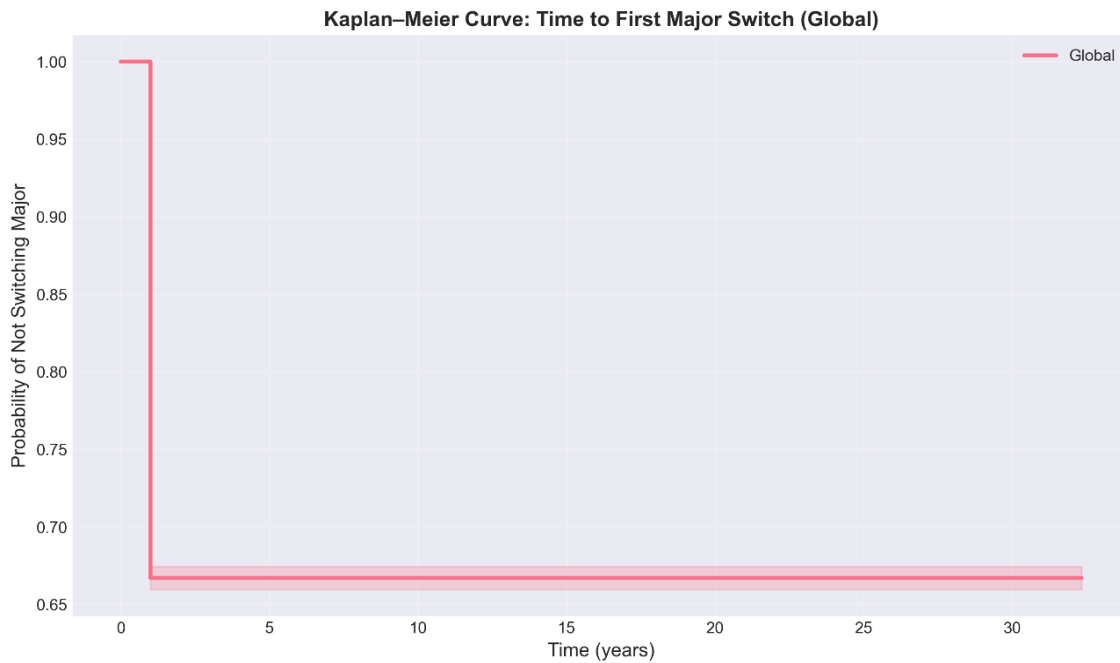

The majority of students (78.6%) never change major and are therefore right-censored in Outcome B. As seen in Figure 5, temporal stratification by entry period shows that switching rates have increased substantially over time, peaking in P3 (2000–2009) at around 28.4%. This rise coincides with an expansion of the programme portfolio and greater curricular differentiation, which both widens the menu of options and normalises programme reorientation as a legitimate academic strategy. Table 3 presents the summary of switching probabilities at different time intervals.

**Figure 5.** Kaplan–Meier survival curves for first major switch by entry period

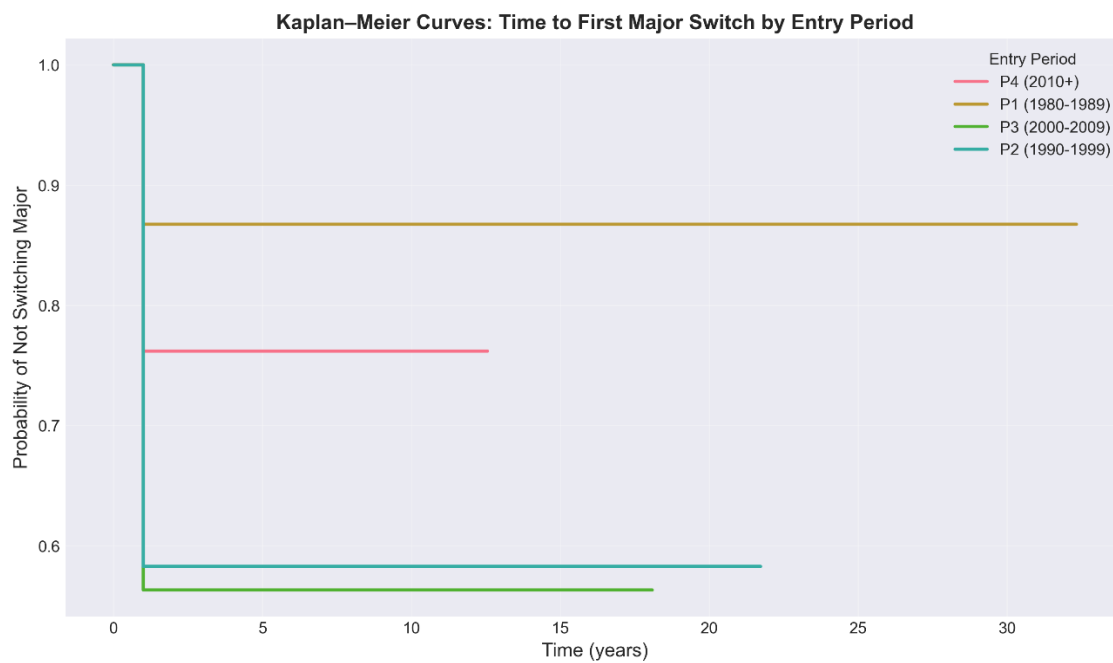

**Table 3.** Summary of switching probabilities at 0.5, 1, 2, and 3 years for Outcome B

| outcome | stratum | stratum_value | n_total | n_events | n_censored | median_survival_time |
|---|---|---|---|---|---|---|
| B | Global | All | 24132 | 5175 | 18957 | inf |
| B | Entry Period | P4 (2010+) | 6346 | 981 | 5365 | inf |
| B | Entry Period | P1 (1980-1989) | 3927 | 322 | 3605 | inf |
| B | Entry Period | P3 (2000-2009) | 9215 | 2614 | 6601 | inf |
| B | Entry Period | P2 (1990-1999) | 4644 | 1258 | 3386 | inf |

These patterns confirm that major switching operates as a mechanism of early academic adjustment rather than as a symptom of long-term disengagement. Students who switch tend to do so quickly, in response to early signals of misalignment between interests, preparation, and programme characteristics, while those who remain in their initial major either progress or ultimately converge into the long-tail dropout regime captured by Outcome A.

**4.4 Comparative interpretation of the two temporal regimes**

The combined analysis of Outcomes A and B reveals two distinct temporal regimes within the academic ecosystem:

- **Early-switching regime (Outcome B).** Programme changes are tightly clustered within the first 12–18 months, suggesting a relatively swift process of exploration, reconsideration, and reallocation of academic goals. This regime is sensitive to portfolio breadth, orientation practices, and the timing of diagnostic assessments.

- **Long-tail dropout regime (Outcome A).** Disengagement extends over many years, with a steep yet sustained decline in survival over a four- to five-year period, followed by a long tail of late dropout. This regime appears shaped by structural friction, repeated course failures, and regulatory mechanisms that allow students to maintain enrolment status despite minimal progression.

Crucially, high early mobility does not translate into high early dropout. Instead, failure is systematically delayed. Horizontal mobility serves as an early adjustment channel, while vertical exit occurs after prolonged periods of stagnation. The system thus combines *adaptive exploration* with *inefficient persistence*, generating the Stagnant Persistence Paradox: students are retained but not advanced, and dropout arrives late, after years of accumulated temporal and economic cost.

## 5. DISCUSSION
### 5.1 The paradox of stagnant persistence
The central finding of this study is the high temporal cost associated with academic failure, captured by the 4.33-year median survival time until definitive dropout. This duration, which in many international contexts would already exceed the nominal length of a bachelor's degree, exposes the Stagnant Persistence Paradox: the academic system retains students for an inordinately long period, characterised by institutional enrolment without sufficient curricular velocity, only to see them ultimately depart. Students are effectively lingering in the system, absorbing institutional resources and incurring substantial personal opportunity costs (wasted time, foregone earnings, delayed entry into the labour market).

In contrast with settings where dropout is concentrated in the first year or two of study (e.g., Murtaugh et al., 1997; Paura & Arhipova, 2019), the long-tail nature of the dropout curve here suggests that attrition is not a single early shock but a protracted process of cumulative friction. Students remain enrolled while repeatedly encountering bottlenecks—core basic science courses, restrictive prerequisite chains, administrative hurdles—which they fail to clear despite extended tenure. Survival, in this sense, becomes a poor proxy for success.

For institutions, this delayed failure represents a major inefficiency. Headcount-based retention statistics can give the illusion of health, as large numbers of students remain on the books, yet a substantial share are effectively trapped in early segments of the curriculum. For students, prolonged persistence without progress can erode motivation, create sunk-cost dilemmas, and exacerbate inequalities when those with fewer external resources persist in unproductive trajectories for longer.

### 5.2 Temporal efficiency, risk, and curricular velocity
The temporal stratification by entry period reveals a critical tension between equity and efficiency. Later cohorts, particularly P4, exhibit longer median survival times and lower observed dropout proportions within the available follow-up. Superficially, this could be read as evidence that the faculty has become more inclusive or better at supporting students from diverse backgrounds. However, longer survival in a context of rigid curricula and limited progression opportunities may instead reflect delayed failure rather than genuine inclusion.

To interpret this tension, we propose reframing institutional key performance indicators away from static retention and towards *curricular velocity*: the ratio between time spent in the system and meaningful advancement through structural milestones. Examples include time-to-clearance of the first-year basic cycle, time-to-accumulation of a given credit threshold, and time-to-first-major-switch when initial allocation is misaligned. In survival terms, this implies monitoring not only whether students remain enrolled but also the hazard of progression versus stagnation at specific points in the curriculum.

By quantifying trajectories in terms of time and risk, institutions can identify pressure points where students spend excessive chronological time with limited academic gain. In the present case, the evidence suggests that a significant proportion of dropouts leave after several years spent repeating early-stage courses, indicating that policies which merely prolong enrolment without restructuring bottlenecks may inadvertently increase the temporal cost of failure.

**5.3 Methodological focus on time-to-event modelling**
Methodologically, this paper demonstrates the value of integrating survival analysis with a trajectory-based reconstruction of student histories. Rather than treating dropout as a static binary label, we define and model two distinct time-to-event outcomes—definitive dropout and first major switch—within a common analytical framework. This allows us to distinguish conceptually and empirically between early adaptive behaviour (switching) and long-term failure (dropout), capturing the coexistence of two temporal regimes under a shared set of institutional rules.

The approach contributes to a growing line of work urging higher education analytics to move beyond descriptive dashboards or opaque predictive models (Arnold & Sclater, 2021; Gašević et al., 2022). By reconstructing trajectories as spells, enforcing strict time-ordering, and isolating baseline features from emergent behaviours, CAPIRE's temporal module aligns with calls for leakage-aware, principle-based use of administrative data. Survival analysis then becomes not merely a statistical technique, but a way of embedding temporal logic into institutional indicators—highlighting when risk is highest, how long non-completion takes, and where interventions might compress risk into less damaging intervals.

Importantly, the non-parametric nature of the models used here keeps the pipeline accessible. Institutions with basic administrative data and modest analytic capacity can replicate Kaplan–Meier estimation and stratification by cohort, deriving substantive insights without immediately requiring high-dimensional modelling or complex machine-learning infrastructure.

**5.4 Implications for institutional metrics and curriculum governance**
The findings suggest several directions for institutional action centred on temporal efficiency:

- **From retention to progression.** University KPIs should move beyond counting how many students remain enrolled to tracking how far they progress relative to time-on-programme. Indicators such as credits-per-year, probability of clearing the first-year curriculum within two chronological years, or time-to-completion of gateway modules provide sharper views of curricular velocity.
- **Early stagnation detection.** Survival curves can be used to identify *drifting students*—those who, by their third chronological year, have not yet completed the first-year curriculum. Policies could mandate early decision points (intensive remediation, structured advising, or guided redirection to alternative programmes) well before the median dropout time, reducing years of unproductive persistence.
- **Temporal audits of reforms.** Because plan changes and regulatory shifts leave identifiable signatures in mobility and survival profiles, institutions can use time-to-event indicators as audit tools for curriculum reform. Increases in median dropout time without corresponding gains in completion may signal that reforms are stretching failure rather than alleviating it (Carvalho et al., 2022; Miller & Torres, 2024).

These implications do not require abandoning traditional retention and completion metrics. Instead, they expand the analytic repertoire to include risk over time, enabling a more nuanced understanding of how structure and regulation shape the temporal cost of education.

**5.5 Limitations and future research**

While the study benefits from a uniquely rich, long-term dataset, several limitations remain. Background variables such as socio-economic status, secondary-school characteristics, and detailed academic performance are incomplete or inconsistently recorded across the forty-year window. This limits the extent to which we can attribute differences in survival patterns to compositional changes versus structural factors. Survival estimates for the most recent cohorts are also constrained by shorter follow-up windows, particularly for late dropout.

Furthermore, the analysis focuses on overall time-to-event, rather than modelling hazards at specific points in the curriculum or for particular subpopulations (e.g., women, first-generation students, or entrants from vocational schools). Future work could integrate socio-demographic and performance data to identify which groups are most exposed to long-tail dropout versus early switching, and to examine whether the Stagnant Persistence Paradox is uniformly distributed or concentrated in specific majors or student profiles.

Finally, this paper employs non-parametric survival methods by design, privileging interpretability and robustness. Extending the framework to semi-parametric or fully parametric models, including competing-risks formulations and multi-state models, would allow more granular estimation of covariate effects and interactions. Nonetheless, the descriptive survival patterns documented here already provide a strong case for rethinking how temporal efficiency is measured in engineering education.

## 6. CONCLUSION

This study analysed forty years of student trajectories in a large engineering faculty through a survival-analytic lens, focusing on two key outcomes: time-to-definitive-dropout and time-to-first-major-switch. By integrating a leakage-aware trajectory reconstruction (spells and typed transitions) with Kaplan–Meier estimation, we characterised the temporal regimes of academic failure and adaptive mobility within a rigid, high-friction educational system.

The results reveal a stark asymmetry between early switching and late dropout. Major switching is concentrated within the first academic year and functions as an early mechanism of academic adjustment, while definitive dropout is a long-tail process with a median time of 4.33 years and a substantial right tail. This configuration gives rise to the Stagnant Persistence Paradox: students are retained in the system for extended periods without commensurate curricular advancement, and institutional indicators that prioritise raw retention risk underestimating the temporal cost of non-completion.

Beyond its substantive findings, the paper offers a transferable blueprint for institutions seeking to incorporate time-to-event analysis into their analytics infrastructures. By shifting attention from static status categories to dynamic measures of curricular velocity, survival analysis can support more informed curriculum design, more targeted advising, and more honest evaluation of equity-efficiency trade-offs. In systems where tuition is free but time is not, quantifying the temporal cost of failure is not merely a technical exercise; it is a prerequisite for responsible governance of mass higher education.

Finally, it is crucial to clarify that the pursuit of 'temporal efficiency' advocated here does not imply a prioritisation of industrial throughput over academic quality or the necessary maturation of learning. Rather, it addresses an ethical imperative rooted in equity. In tuition-free systems, the primary cost of education is the student's time—a non-renewable resource that weighs most heavily on those from less privileged backgrounds. By highlighting the paradox of stagnant persistence, we argue that a system which retains students for years without meaningful progression is not 'inclusive'; it is structurally deceptive. Therefore, metrics of curricular velocity should be understood as tools to protect students from the high opportunity costs of delayed failure, ensuring that university tenure translates into either timely success or early, constructive redirection.